\documentclass[12pt]{elsart}
\usepackage{graphicx, amssymb}
\journal{Physics Letters B}
\begin{document}

\begin{frontmatter}

\title{Study of system- size effects in multi- fragmentation
using Quantum Molecular Dynamics model}
\author[puchd]{Jaivir Singh}, 
\author[puchd]{Rajeev K. Puri \corauthref{cor}},
\ead{rkpuri@pu.ac.in}
\author[france]{J$\ddot{o}$rg Aichelin}

\corauth[cor]{Corresponding Author.}
%\author{J$\ddot{o}$rg Aichelin}}

\address[puchd]{Department of Physics, Panjab University, Chandigarh -160 014, India.}
\address[france]{SUBATECH, Ecole des Mines de Nantes, 4 rue Alfred Kastler, F-44070 Nantes Cedex, France.}
%%%%%%%%%%%%%%%%%%%%%%%%%%%%%%%%%%%%%
\begin{abstract}
We report, for the first time, the dependence of the multiplicity of
different fragments on the system size employing a quantum molecular dynamics
model. This dependence is extracted from  the simulations of symmetric
collisions of Ca+Ca, Ni+Ni, Nb+Nb, Xe+Xe, Er+Er, Au+Au and U+U at
incident energies between 50 A MeV and 1 A GeV. We find that
the multiplicity of different fragments scales with the size of the system which can be
parameterized by a simple power law.
\end{abstract}
%%%%%%%%%%%%%%%%%%%%%%%%%%%%%%%%%%%%%%%%%%%%%%%
\end{frontmatter}

%%%%%%%%%%%%%%%%%%%%%%%%%%%%%%%%%%%%%%%%%%
%\section{introduction}
The breaking of colliding nuclei into dozen of fragments has been a central topic both in experimental and theoretical Nuclear Physics
 research. The
theoretical interests have been concentrated on the understanding of the
reaction
mechanism behind fragmentation. One has also tempted to look for the role
of in-medium cross section and  nuclear equation of state in fragmentation.
These studies have been carried out  by analyzing the fragment multiplicity and their
kinetic energy spectra using heavy target/projectile [1-4,6].
However, a very little attention  has been paid
to the system size effects in multi-fragmentation although it may
not only explore the effect of ratio of surface to volume in
fragmentation, but can also
throw the light on the dynamical effects which are expected to
 increase with the increasing size of the system.
%%%%%%%%%%%%%%%%%%%%%%%%

Recently, the FOPI collaboration \cite{Reis99} analyzed the  system size effects
in fragmentation by
studying the reactions of Ni+Ni, Ru+Ru, Xe+CsI and Au+Au. They analyzed the
spectra by dividing the interacting matter into
spectators and participants. The spectator fragments show
the well known universality \cite{Schu96} whereas the
intermediate mass fragments emitted from the participant source scale 
with the size of the emitting source indicating  the role of
the expansion of the matter in multi-fragmentation. The role of the
expansion has already been  established in collective flow [7].\\

%%%%%%%%%%%%%%%%%%%%%%%%%%%%%%%%%%%%
Motivated by these findings, we present here, for the first time, a complete
study of the dependence of the fragment multiplicity on the size of the colliding
(symmetric) nuclei by studying the reactions of different nuclei with masses
ranging between 40 and 238. We shall
show that the multiplicity of all kinds of fragments scales with the
size of the system which can be parameterized by a simple power law $\propto
A_{tot}^\tau (A_{tot}$ is the mass of the composite system) at all incident
energies and impact parameters .\\
%%%%%%%%%%%%%%%%%%%%%%%

A word of caution should be added here: It has been shown and discussed
extensively in the literature that the mass yield curve approximately
obeys a power law behavior $ \propto A^{-\tau}$ [2,4]. It  has been
conjectured  (though controversial) that this  behavior is an indication
of the phase transition
between a gaseous and liquid phase of the nuclear matter. This power law
behavior obtained for  the mass or charge distribution  is for a "{\bf
given system}".
The power law dependence, which  we are talking about, is something very different.
We shall show that the  multiplicity of a given fragment
(or a group of fragments) scales with the size of the interacting system which
can be parameterized in terms of a power law function.\\

%%%%%%%%%%%%%%%%%%%%%%%%%%%%
The basis of our investigation is a quantum molecular dynamical (QMD) model
\cite{Aich91}, where nucleons interact via two and three- bodies which
preserves the particle correlation and fluctuations. Here, each nucleon is
represented by a Gaussian wave packet with width $\sqrt{L}$ centered around
the mean position $\vec r_i(t)$ and the mean momentum $\vec p_i(t)$:

\begin{equation}
\phi_i(\vec{r}, \vec{p}, t)~=~\frac{1}{(2\pi L)^{3/4}}~e^{\left[
-(\vec{r}~-~\vec{r}_i (t))^2 /4L \right]}~~
e^{\left[ i{\vec p}_i (t) \cdot
\vec{r}/\hbar \right]}.
\label{e1}
\end{equation}

The Wigner distribution of a system with ($A_T~+~A_P$) nucleons is given by
\begin{equation}
f(\vec{r},\vec{p},t)~=~\sum_{i=1}^{A_T+A_P} \frac{1}{(\pi \hbar)^3}
e^{\left[ -(\vec{r}~-~\vec{r}_i (t))^2/2L \right]}~~e^{\left[
-(\vec p~-\vec p_i (t))^2~2~L/{\hbar}^2 \right]}.
\end{equation}
We shall use here a  static soft equation of state along with  energy
dependent nucleon- nucleon cross section. The detail discussion about
the different equations of state and cross sections can be found in ref.\cite{Aich91}.
 We follow the nucleons till 300 fm/c and then freeze their coordinates.
 This
 time is long enough to assume that the reaction has finished. The
 frozen phase space of nucleons is,  then, clusterized to obtain the fragments.
 The clusterization is performed within Minimum Spanning Tree [MST] method
 which binds two nucleons in a fragment if they are closer than 4 fm.\\
%%%%%%%%%%%%%%%%%%%%%

Here we simulate several thousand events involving the different symmetric
colliding nuclei like $^{40}Ca + ^{40}Ca, ^{58}Ni + ^{58}Ni, ^{93}Nb + ^{93}Nb,
^{131}Xe + ^{131}Xe, ^{168}Er + ^{168}Er, ^{197}Au + ^{197}Au $~and~ $^{238}U + ^{238}U$
at incident energies between 50A MeV and 1A GeV and at different impact
parameters $\hat{b} = b/b_{max}$; $b_{max} = R_1+R_2$, $R_i$ is the radius of
target/projectile.  By varying the size of the symmetric system,
 the system size effect can be studied without varying the
asymmetry  and  excitation energy of the reaction. Note that the  ALADIN
experiment varies the asymmetry of a reaction \cite{Schu96}
whereas the FOPI experiments are for symmetric nuclei \cite{Reis99}.
The wide range of incident energy between 50 A MeV and 1 A GeV
gives us possibility to study the multi-fragmentation in regions
where different reaction mechanisms have been proposed [6].\\
%%%%%%%%%%%%%%%

In Fig. 1, we display the largest remnant, the number of emitted nucleons
and light fragments with mass = 2 at 50A MeV and 400A MeV. The
multiplicity of the light mass fragments (LMF's)
$2 \le A \le 4$, the medium mass fragments (MMF's) $3 \le A \le 14$ and
the intermediate mass fragment (IMF's) $5 \le A \le $min$[A_{tot}/6, 65]$
is displayed in fig.2. Note that the MMF's and IMF's excludes the heaviest
fragment.
%%%%%%%%%%%%%%%%%%%%%%
\\

The general behavior of all fragments follows the well known trends.
In peripheral collisions ($\hat{b} = 0.6$), the largest fragment
$A^{max}$ follows roughly the reaction geometry insofar the non overlapping nucleons for
the largest remnant.  The size of the heaviest fragment scales with
the size of the spectator  matter which depends on the size of the
interacting system. The light charge particles ( i.e. the emitted
nucleons, the fragment with mass= 2 and LMF's) follow the same trend at all
incident energies. Their multiplicity is maximum at $\hat{b}$ = 0 and
which is  followed
by $\hat{b}$ = 0.3 and 0.6. The emission of the light charge
particles depends on the size of the participants which decreases with the
decrease in the overlapping volume.  One also sees that the
number of light charge particles emitted at 50 A MeV is much smaller than
at 400 A MeV.  Most of the   matter at higher incident
energies ( i.e. at 1  A GeV)  is in the form of light
charge particles. 
At higher incident energies, the nucleon-nucleon
collisions (with large momentum transfer) destroy the correlation in the
participant matter and only light particles survive from the reaction
zone. At lower incident energies, most of the collisions are Pauli
blocked and therefore, many nucleons in the reaction zone survive the
reaction without suffering collisions with a large momentum transfer.
The energy received by a target in peripheral collisions is not enough
to excite the matter far above the Fermi level, therefore, a heavier
largest fragment $A^{max}$, and small number of nucleons/LMF's and
MMF's are emitted at peripheral collisions ($ \hat{b} = 0.6$).
We also see that the maximum production of medium mass and intermediate
mass fragments
occurs for $\hat{b}$=0 at 50 A MeV which shifts towards $\hat{b}$ = 0.3
at 400 A MeV \cite{Schu96,Aich91}. With increasing incident
energy, the multi-fragmentation becomes more and more a phenomenon of
peripheral reaction. Due to larger overlap  between  the colliding nuclei
at $\hat{b}$ = 0,  the excitation in the system at 400 A MeV  does not
allow the formation of heavier fragments, therefore, we see small
number of MMF's and IMF's.  The situation improves at $\hat{b}$ = 0.3.
 At peripheral collisions ($\hat{b} $= 0.6), the amount of excitation,
 compression and nucleon-nucleon collisions decreases and therefore,
  heavier residual  fragment survives.\\

The most interesting outcome of figs. 1 and 2 is that in all cases,
independent of the mass of the fragment, incident energy ( and excitation energy)
and impact parameter, the heaviest fragment and the multiplicity of all
kind of fragments ( i.e. of  nucleons, light, medium and intermediate
mass fragments) scales with the size  of the system which can be nicely
parameterized by a power law = $c\cdot A_{tot}^\tau; A_{tot}$ is the
mass of composite system. The values of constants $ c $ and $\tau$ depend on the
 size of the fragment and on the incident energy and impact parameter
of the reaction. The dependence of
the power factor $\tau$ on incident energy and impact parameter
is displayed in figure 3.
We also tried 
a function form  C$\cdot e^{-\tau A}$, but fits
were worse than obtained with a power law.
It is worth mentioning that a similar
power law dependence is also obtained for other groups of fragments
like $5 \le A \le 25$.
The power law was also obtained at other incident energies 
E= 100, 200, 600 and 1000A MeV indicating the universality of the power law
behavior for system size effect in multi-fragmentation/clusterization.
\\

In fig.3, we plot the  value of $\tau$  as a function of the
reduced impact parameter
$\hat{b}$ at incident energies 50 and 400 A MeV.  The different panels of the
figures are, respectively,  for heaviest fragments $A^{max}$, free nucleons,  fragments
with mass 2, LMF's, MMF's and IMF's.  We see that the value of
the $\tau$  in most of the cases depends weakly on  the impact parameter. The value of
$\tau$ increases with the incident energy and then saturates for very high energies.
It is close to
2/3  at 50 A MeV whereas it is $\approx 1$  at 400 A MeV.  The
only exception in this trend is the value of $\tau$ (= 2.6)
for IMF's at 400 A MeV and $\hat{b}$ = 0. This  exception is rather
linked with the mass range used to define an IMF.  The lower mass limit of
an IMF is 5 units and from fig. 1(b), we see that the $A^{max}$
in this  reaction fluctuates  around 5 for heavier colliding nuclei
which may lead to the emission of few
intermediate mass fragments. These
fragments cause a large value of $\tau$.                          \\

We also notice that unlike the disappearance of flow (where the energy of
vanishing flow varies as $A^{-1/3}$) \cite{West93}, no unique dependence on
$\tau$ exists.  This behavior of $\tau$  demonstrates that the degree
of clusterization is more for heavier colliding nuclei which could be attributed
to the fast decompression of the system in heavier nuclei at central collisions.
Similar trends are also reported by the FOPI collaboration \cite{Reis99}.
If one plots the reported results of FOPI experiments \cite{Reis99} as a function
of size of the system, a similar power- law fit can also be obtained \cite{Reis99}.
Here one should keep in the mind that the analysis of FOPI experiments is done
for participant region only. Our present result includes both the participant and
spectator regions. It is worth mentioning that the value of $\tau$ may
 depend on the clusterization model one is using. As has been noted in Ref.(3),
  this  scaling of
the multiplicity of fragments with  total mass of the system, draws
two important aspects: (i) the role of the surface to volume ratio in
clusterization and second, the dominance of the expansion following the compression
over the competition between the attractive mean field and repulsive
Coulomb force in multi-fragmentation. The increase in the multiplicity of
the fragments with mass of the system points toward the fact that the expansion
of the compressed matter is much faster in heavier systems compared to
lighter systems.\\
%%%%%%%%%%%%%%%%%%%%%

Summarizing, using quantum molecular dynamics (QMD) model, we find, for
the first time, a power- law behavior for the system- size effect in
fragmentation at all incident energies between 50 A MeV and
1 A GeV and for all colliding geometries.  This dependence exists
for any kind of fragment/group of fragments and can be  parameterized
in terms of  a power law. \\
%%%%%%%%%%%%%%%%%%%%%%%%%%%%%%%%%%%
{\it This work was supported by Council of Scientific and Industrial
Research grant No. 03 (0823) /97/EMR-II.}

%%%%%%%%%%%%%%%%%%%%%%%%%%%%%%%%%%%

%%%%%%%%%%%%%%%%%%%%%%%%%%%%%%%%%%
\newpage

\begin{figure}
\begin{center}
\includegraphics*[width=5cm]{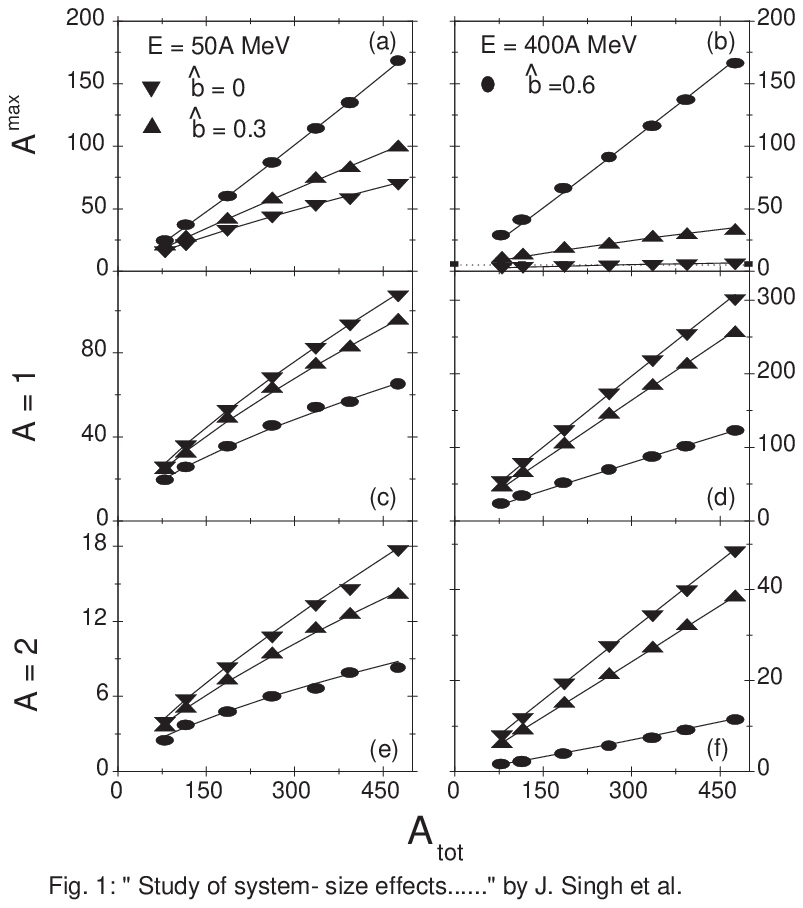}
\end{center}
\caption{The heaviest fragment $A^{max}$, the multiplicity of
free- nucleons and fragments with mass A = 2 as a function of composite
mass of the system $A_{tot} (= A_T + A_P)$ at different scaled impact
parameters $\hat{b}$. The left hand side is at 50A MeV whereas right
hand side is at 400A MeV. The displayed results are at 300 fm/c.}
\end{figure}

\begin{figure}
\begin{center}
\includegraphics*[width=5cm]{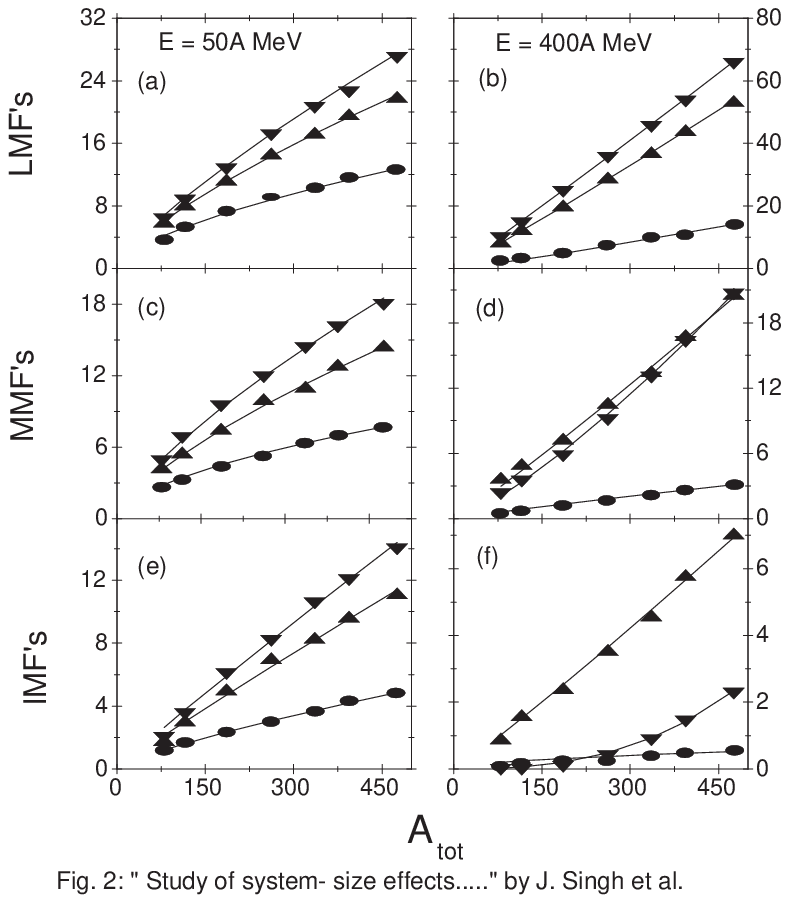}
\end{center}
\caption{Same as fig. 1, but for light mass fragments (LMF's)
($2 \le A \le 4$), medium mass fragments (MMF's) ($3 \le A \le14$) and
intermediate mass fragments (IMF's) ($5 \le A \le min [1/6A_{tot},65]$).}
\end{figure}

\begin{figure}
\begin{center}
\includegraphics*[width=5cm]{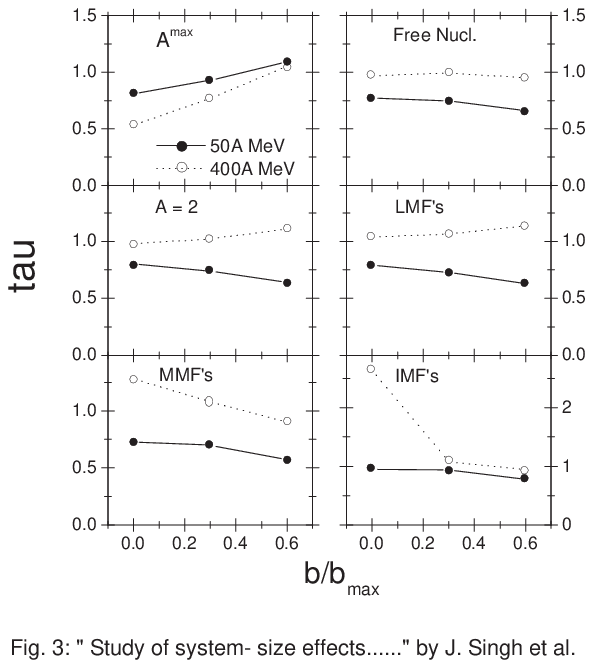}
\end{center}
\caption{The $\tau$  as a function of the reduced impact parameter
$\hat{b}$. The solid and dashed lines shows, respectively, the
results at 50 A MeV and 400 A MeV.}
\end{figure}

\end{document}